\documentclass[[11 pt, a4paper, table]{article}
\usepackage[latin9]{inputenc}
\usepackage{amsmath}
\usepackage{mathtools}
\usepackage[unicode=true,pdfusetitle,
 bookmarks=true,bookmarksnumbered=false,bookmarksopen=false,
 breaklinks=false,pdfborder={0 0 1},backref=false,colorlinks=false]
 {hyperref}

\makeatletter

\DeclareTextSymbolDefault{\textquotedbl}{T1}

\@ifundefined{date}{}{\date{}}
\usepackage{tikz}
 \usepackage{stmaryrd}
  \usepackage{stackrel}

\usepackage{subfig}
\usepackage{rotating}
\usepackage{tabularx}
\usepackage{graphicx}
\usepackage{afterpage}
\usepackage{array}
\usepackage{multirow}

\usepackage{latexsym}
\newcommand{\be}{\begin{equation}}
\newcommand{\ee}{\end{equation}}
\newcommand{\bea}{\begin{eqnarray}}
\newcommand{\eea}{\end{eqnarray}}
\newcommand{\beas}{\begin{eqnarray*}}
\newcommand{\eeas}{\end{eqnarray*}}

\def\tr{{\rm Tr}}

\def\XXint#1#2#3{{\setbox0=\hbox{$#1{#2#3}{\int}$ }
\vcenter{\hbox{$#2#3$ }}\kern-.5\wd0}}

\makeatother

\begin{document}
\title{"Glueballs" in the Quantum Mechanics of three large massless Yang-Mills coupled Matrices}
\author{
Jo\~ao P. Rodrigues\thanks{Email: joao.rodrigues@wits.ac.za} 
\\
 \\
 National Institute for Theoretical and Computational Sciences\\
 School of Physics and Mandelstam Institute for Theoretical Physics
\\
 University of the Witwatersrand, Johannesburg\\
 Wits 2050, South Africa \\
 }
\maketitle
\begin{abstract}
We use a loop truncated Jevicki-Sakita effective collective field Hamiltonian to obtain, over a very large range of values of 't Hooft's coupling, and directly in the large $N$ limit, the large $N$ (planar) ground state energy, the planar ground state expectation values of invariant correlators, and the $1/N$ spectrum of the quantum mechanical system of three massless Yang-Mills coupled matrices. This captures the dynamics of the (residual) gauge invariant sector of the spatially reduced $3+1$ dimensional pure Yang-Mills theory, in the large $N$ limit.The large $N$ loop space constraints are handled by the use of master variables. As is the case for two matrices, the method is highly efficient directly in the massless limit, and it reproduces to a very high precision the scaling dependence of physical quantities, determined by their dimensions, on the dimensionful 't Hooft coupling. We obtain the bound state masses of "glueballs", their quantum numbers and ensuing degeneracies.   
 \end{abstract}

\section{Introduction}
There is recent interest in the application of numerical methods to a direct study of the large $N$ limit \cite{tHooft:1973alw} properties of multi-matrix systems \cite{Anderson:2016rcw, Lin:2020mme, Han:2020bkb, Kazakov:2021lel, Koch:2021yeb, Kazakov:2022xuh}
\footnote{We have in mind the path integral or the quantum mechanics of a finite number of hermitian or unitary matrices. For more recent related work, see, e.g., \cite{Kazakov:2024ool, Li:2024wrd,Li:2024ggr, Cho:2024kxn, Jevicki:2025ybt}.}.  This is not surprising, given their importance as
providing reduced ansatzes for large $N$ gauge theories, both in the path integral 
\cite{Eguchi:1982nm, Bhanot:1982sh, Parisi:1982gp, Gross:1982at, Das:1982ux} and Hamiltonian formulations \cite{Neuberger:1982ne, Kitazawa:1982wn}, and, since their interpretation as $D0$ branes \cite{Polchinski:1995mt}, as possible candidates for a non-perturbative definition of M theory\cite{Banks:1996vh}, also argued to be valid for the path integral \cite{Ishibashi:1996xs}.  The $AdS/CFT$ correspondence \cite{Maldacena:1997re, Gubser:1998bc, Witten:1998qj} has highlighted the importance of $\mathcal{N}=4 $ SYM theory, with its bosonic adjoint scalar sector, and ensuing integrability properties \cite{Beisert:2010jr} and Hamiltonian reductions \cite{Berenstein:2002jq},  \cite{deMelloKoch:2002nq, Beisert:2002ff, Kim:2003rza}. They are used in the study of black holes \cite{Kazakov:2000pm, Cotler:2016fpe, Maldacena:2023acv}\footnote{There is a vast literature on matrix models; we have tried to highlight only some key developments in their application, with emphasis on YM coupled systems}.

Whether one bootstraps by iterating loop equations and refining initial guesses \cite{Lin:2020mme} or by semidefinite programming \cite{Anderson:2016rcw, Han:2020bkb, Kazakov:2021lel, Kazakov:2022xuh}, or whether one uses optimization with respect to master variables \cite{Koch:2021yeb}, all these approaches have to take into account the existence of non-negative loop constraints, long identified in  \cite{Jevicki:1982jj, Jevicki:1983wu, Jevicki:1983hb, Rodrigues:1985aq}.

In this communication, we study the large $N$ properties of the gauge invariant sector of the Hamiltonian of three massless hermitian matrices interacting via a Yang-Mills potential. This system is the spatially reduced $3+1$ dimensional pure Yang-Mills theory in the large $N$ limit, and for physical states satisfying Gauss' law. As is the case for two massless Yang-Mills coupled hermitian matrices \cite{Mathaba:2023non}, we are able to study the system directly in the large $N$ and massless limits. In general \cite{Morita:2020liy, Pateloudis:2022ijr}, properties of the massless system are obtained by extrapolation of a system with a finite mass parameter to zero. In our case, being able to work directly in the massless limit, we obtain and confirm the expected scaling dependence of energies and planar correlators on the coupling constant, to a high degree of precision, and over a range of 't Hooft couplings close to four orders of magnitude.  . 

Our approach is based on the collective field theory hamiltonian of Jevicki and Sakita \cite{Jevicki:1979mb}. This Hamiltonian is an exact re-writing of a given theory in terms of its (gauge) invariant variables. The large $N$ (planar) background is then obtained semiclassically as the minimum of an effective potential $V_{eff}$ and, when expanded about this large $N$ background, the collective field theory Hamiltonian generates $1/N$ corrections systematically \footnote{For a single matrix based example, see for instance   \cite{Das:1990kaa},    \cite{Demeterfi:1991cw}      }
\footnote{We consider matrix valued systems in this communication, but the same is true of vector valued field theories}.

For the numerical results, we use a truncated collective field hamiltonian. The issue of constraints is addressed by the use of "master variables"  \cite{Jevicki:1983wu, Koch:2021yeb,Jevicki:1983hb}. These are variables that satisfy the constraints explicitly, of which the original variables are an example. For three matrix systems, we keep one of the matrices diagonal and the other two as arbitrary $N \times N$ hermitian, so that there are $N(2N+1)$ master variables. 
The effective potential is then minimized with respect to these variables, from which the planar large $N$ energy and planar expectation values of invariant correlators are obtained. In addition to satisfying the constraints in the planar limit, master variables can be used to set up the spectrum equations of the theory \cite{Jevicki:1983hb}.

As a reduced $3+1$ dimensional Yang-Mills system,  the study of the spectrum and ensuing presence of mass gaps, scaling behaviour, quantum numbers and degeneracies is of particular importance (interest?). We believe that currently, this is the only method able to provide information about the large $N$ spectrum of the theory. 

This article is organized as follows: after the current Introduction, Section $2$ briefly describes the method, the loop truncation and the use of master variables in dealing with the loop space constraints and in obtaining both the large $N$ energy and background and the $1/N$ spectrum. A more detailed description can be found in \cite{Mathaba:2023non} and \cite{Koch:2021yeb}, In Section $3$ we apply the method to the large $N$ limit of the quantum mechanics of three massless Yang-Mills coupled matrices. The method displays perfect stable convergence directly in this massless limit, with physical quantities exhibiting the scaling behaviour determined by their dimensions to a very high level of precision, both for planar quantities and for the $1/N$ spectrum. In other words, the loop truncated collective field Hamiltonian is entirely consistent with the scaling properties of the full massless theory.  For the spectrum, bound "glueball" states develop well defined mass-gaps. Their degeneracies and quantum numbers are identified. In Section $4$, we compare results of the large $N$ spectrum of the reduced model to those of lattice gauge theories,  and present a brief discussion and outlook. 

\section{Method and loop truncation}

\hspace{10pt} 
We consider the quantum mechanics of three $N \times N$ hermitian matrices $X_A, \, A=1,2,3$, interacting via a Yang-Mills potential:
\begin{equation}\label{FreeHam}
\hat{H} = \frac{1}{2}  \sum_{A=1}^{3} \tr{P_A^2} +  \frac{m^2}{2} \sum_{A=1}^{3} \tr{X_A^2}  - \frac{g_{YM}^2}{2 N} \sum_{A\ne B}^3 \tr [X_A,X_B]^2 =  \frac{1}{2}  \sum_{A=1}^{3} \tr{P_A^2} + \tr (V(X_A)).\end{equation}
$P_A$ is canonical conjugate to $X_A$, and $m$ is a mass. We will only consider the massless case $m=0$ in this communication. Note that in terms of our conventions,  't Hooft's coupling $\lambda$ is $\lambda= g_{YM}^2$.

The $U(N)$ invariant loops are single traces of products of the matrices $X_A$, up to cyclic permutations:
\begin{equation*}
\phi(C) = \tr (... X_1^{m_1} X_2^{m_2}  X_3^{m_3} X_1^{n_1} X_2^{n_2} X_3^{n_3} ...)\, .
\end{equation*}
For instance, with two matrices one has $[1\, 1]= \tr(X_1^2)\, , [1 \,2]=\tr(X_1 X_2)\, , [1 \,3]=\tr(X_1 X_3)\, , [2\, 2]= \tr(X_2^2) \, ,  [2 \,3]=\tr(X_2 X_3) \, ,  [3\, 3]= \tr(X_3^2)$, with three matrices $[1 \,1 \,1] = \tr(X_1^3)\, , [1\, 1 \,2] = \tr(X_1^2 X_2)\, , [1\, 1 \,3] = \tr(X_1^2 X_3)\ ,[1\, 2 \,2] = \tr(X_1 X_2^2)$, $[1\, 2 \,3] = \tr(X_1 X_2 X_3)$, etc., with an obvious notation. We will continue to refer to the invariant variables as ``loops", for historical reasons. 

The collective field Hamiltonian \cite{Jevicki:1979mb} in terms of the invariant loops $\phi(C)$ takes the form 
\begin{equation*}
H'_{col} = \frac{1}{2} \sum_{C,C'}  \pi^{\dagger}(C) \Omega(C,C') \pi(C') + \frac{1}{8} \sum_{C,C'} w(C) \Omega^{-1}(C,C') w^{\dagger}(C') + V(\phi) + \Delta H' ,
\end{equation*}
where $\pi(C)$ is the canonical conjugate to $\phi(C)$, and  
\begin{eqnarray*}
\Omega(C,C') &=&\sum_{A=1}^3 \tr \left( \frac{\partial \phi^{\dagger}(C) }{\partial X^{\dagger}_A} \frac{\partial \phi(C') }{\partial X_A} \right) = \sum_{C"} y(C,C',C'') \phi(C'') \\
w(C) &=& \sum_{A=1}^3 \tr \left(  \frac{\partial^2 \phi(C) }{\partial X^{\dagger}_A\partial X_A} \right)  = \sum_{C',C"} z(C,C',C'') \phi(C')  \phi(C'')  .  \end{eqnarray*}
$\Omega(C,C')$ joins a loop $C$ of length (number of matrices in the loop) $l(C)$ and another of length $l(C')$ into a number of loops of length $l(C)+l(C')-2$. $w(C)$ splits the loop $C$ of length $l(C)$ into sets of two loops $C'$ and $C''$ with total lengths $l(C)-2$.
$\Delta H'$ contains subleading (in $1/N$) counterterms that need not be considered for the large N background and the spectrum.


In order to exhibit explicitly the large $N$ dependence, we let 
\begin{equation*}
\phi(C) \to \frac{\phi(C)}{N^{\frac{l(C)}{2}+1}} = \frac{\tr (... X_1^{m_1} X_2^{m_2} X_1^{n_1} X_2^{n_2}...)}{N^{\frac{l(C)}{2}+1}} , \hspace{6pt} \pi(C) \to N^{\frac{l(C)}{2}+1} \pi(C) \end{equation*}
and obtain
\begin{align}
H_{col} &= \frac{1}{2N^2} \sum_{C,C'} \pi^{\dagger}(C) \Omega(C,C') \pi(C') + N^2 V_{eff} (\phi) \, , \label{Hcoll}\\
V_{eff} (\phi) &\equiv \frac{1}{8} \sum_{C,C'} w(C) \Omega^{-1}(C,C') w^{\dagger}(C') + V(\phi) . \label{Veff}
\end{align} 

It follows that the large $N$ background is the minimum of $V_{eff}$ subject to the constraint that $\Omega(C,C')$ is semi-positive definite.\footnote{The discussion next in this section follows closely that of \cite{Mathaba:2023non} and \cite{Koch:2021yeb}, which are based on \cite{Jevicki:1982jj,Jevicki:1983wu,Jevicki:1983hb}}
\subsection{Truncation of loop space}
For a given $l$ ($l \ge 4$), $\Omega$ is truncated to be a $N_{\Omega} \times N_{\Omega}$ matrix, where $N_{\Omega}$ is the number of loops of length $l$ or less. $\Omega$ itself, however, depends on loops with lengths up to $l_{\rm max}=2 l -2$. If $N_{\rm loops}$ is the number of loops with length $l_{\rm max}$ or less, then it is seen that $V_{eff}$  in (\ref{Veff}) is a function of $N_{\rm loops}$:
\begin{equation*}
V^{trunc}_{eff} (\phi(C), C=1,...,N_{\rm loops}) = \frac{1}{8} \sum_{C,C' = 1}^{N_{\Omega}} w(C) \Omega^{-1}(C,C') w^{\dagger}(C') + V(\phi) \end{equation*}
The following table displays how these numbers grow:
\begin{table}[h]
\begin{center}
\begin{tabular}{||c|c|c||} 
\hline
$l_{\rm max}$& $N_\Omega$  & $N_{\rm loops}$ \\ [0.5ex] 
\hline
6 & 44& 225\\ 
\hline
8 & 75 &1374\\ 
\hline
10 & 225 &9503\\ 
\hline
12 & 540 &69978\\
 \hline
\end{tabular}
\caption{Truncating loop space}
\label{table:1}
\end{center}
\end{table}

\subsection{Planar limit.}
In order to minimize $V^{trunc}_{eff}$ subject to the constraint $ \Omega(C,C') \succeq 0$, we introduce master variables $\phi_{\alpha}$ that explicitly satisfy this constraint:
\begin{equation*}
\Omega(C,C') =\sum_{\alpha} \frac{\partial \phi^{\dagger}(C) }{\partial \phi_{\alpha}} \frac{\partial \phi(C') }{\partial \phi_{\alpha}}\succeq 0 \, .\end{equation*}
Specifically, we choose $X_1$ to be diagonal and $X_2\, , X_3$ arbitrary $N \times N$ hermitian matrices. The master field then has $2N^2+N$ real components $\phi_{\alpha},\, \alpha=1,2,...,N(2N+1).$ 

The planar limit is obtained by minimizing $V^{trunc}_{eff}$ with respect to the master variables. More precisely, at the minimum, 
\begin{align}
\frac{\partial V^{trunc}_{eff}}{\partial \phi_{\alpha}} \equiv \sum_{C=1}^{N_{\rm loops}}\frac{\partial V^{trunc}_{eff}}{\partial \phi(C)}  \frac{\partial \phi(C) }{\partial \phi_{\alpha}} \Big|_{\phi_{\alpha}^0} &= 0 , \,\, \alpha = 1,2,..., N(2N+1) \label{Mmin}\\
\phi_{\rm planar}(C)&\equiv \phi(C)|_{\phi_{\alpha}^0}  , \,\,  C=1,..., N_{\rm loops}.\end{align}
In general, $\partial V^{trunc}_{eff}   / \partial \phi(C) \ne 0$. The planar background is specified by the large $N$ expectation values $\phi_{\rm planar}(C)=\phi(C)|_{\phi_{\alpha}^0}$ of all gauge invariant operators. 
 
The numerical algorithm generalizes in a straightforward way that of two matrices, \cite{Koch:2021yeb} and \cite{Mathaba:2023non}. In this communication, we have chosen a truncation with $l_{\rm max} = 10$, that is, $9503$ $N_{\rm loops}$ and a $225 \times 225$ $\Omega$ matrix. For the master field, we took $N=69$, corresponding to $9591$ master variables. The convergence criteria requires the magnitude of the components of the gradient vector to be of the order of $10^{-10}$.

\subsection{Spectrum}

The $1/N$ expansion is an expansion in terms of loop variables. As such, one lets 
\begin{equation*}
\phi(C)=\phi_{\rm planar}(C) + \frac{1}{N} \eta(C) , \hspace{5pt}  \pi(C)=N p(C)   ,    \end{equation*}
and expands (\ref{Hcoll}) up to second order. Use of master variables ensures that the linear term in $\eta$ vanishes, and from the study of quadratic small fluctuations, one obtains (\cite{Koch:2021yeb} and \cite{Mathaba:2023non}, based on \cite{Jevicki:1983hb}):

\begin{equation}\label{eigSp}
\epsilon_n = \left[ \text{eig}_n\left(  \sum_{C'=1}^{N_{\rm loops}}  \hat{\Omega}_0(C,C') V_0^{(2)} (C',C'') \right) \right]^{1/2}\, ,
\hspace{8pt} V_0^{(2)} (C,C') \equiv \frac{\partial^2 V^{trunc}_{eff}}{\partial \phi(C) \phi^{\dagger}(C')} \Big|_{\phi_{\alpha}^0}
\end{equation}

Note that $\hat{\Omega}_0$ is \underline{not} the same as $\Omega$, but a matrix of size $N_{\rm loops} \times N_{\rm loops}$! In practice, it cannot be calculated in loop space as a loop joining matrix, but at the minimum, it can be obtained from the planar master field $\phi^0_{\alpha}$ as:
\begin{equation*}
 \hat{\Omega}_0(C,C') =\sum_{A=1}^3 \sum_{a,b=1}^{N}\left( \frac{\partial \phi^{\dagger}(C) }{\partial (X^{\dagger}_A)_{ab}} \right)\Bigg|_{\phi^0_{\alpha}} \left( \frac{\partial \phi(C') }{\partial {(X_A)_ {ba}}} \right) \Bigg|_{\phi^0_{\alpha}}, \, C,C'=1,...,N_{\rm loops}\end{equation*}

As a result of the difference in dimensions between $ \hat{\Omega}_0$ and $\Omega_0$, there are $N_{\Omega}$ physical, and in general finite, eigenvalues, with $N_{\rm loops}-N_{\Omega}$ zero eigenvalues \cite{Koch:2021yeb, Jevicki:1983hb}

\section{Massless quantum mechanical system, or reduced large $N$ Yang-Mills theory} 

We study the three matrix Hamiltonian (\ref{FreeHam}) in the massless case.
This system has one dimensional parameter only, $g_{YM}$\footnote{Recall that in terms of our conventions, 't Hooft's coupling $\lambda=g_{YM}^2$.}. Its dimension, that of $\lambda$ and of the fields $X_A$ are:
\begin{equation*}
[g_{YM}]= \frac{3}{2}\, , \hspace{12pt}      [\lambda]= 3\, ,    \hspace{12pt} [X_1]=[X_2]=-\frac{1}{2}\, .
\end{equation*}
 As such, we expect a simple algebraic dependence on $\lambda$ of all physical quantities, simply determined by their dimensions. For instance,
 \begin{equation*}
e = \Lambda_e \, {\lambda}^{1/3} \, , \hspace{7pt}  \tr X_1^2  = \Lambda_{[1 1]}  \, {\lambda}^{-1/3} \, ,  \hspace{7pt}  \tr X_1^4  = \Lambda_{[1 1 1 1]}  \, {\lambda}^{-2/3} \, , \hspace{5pt} \rm{etc.},\end{equation*}
where $e$ is any energy of the system.

We considered $10$ values of $\lambda$, ranging from $e^{-4}$ to $e^{5}$ , chosen to be equally distributed over a logarithmic scale and over almost four orders of magnitude, as shown in Table \ref{table:2}:

\begin{table}[h!]
\begin{center}
\resizebox{\textwidth}{!}
{%
\begin{tabular}{||c|c|c|c|c|c|c|c|c|c||} 
\hline
 \multicolumn{10}{||c||}{$\lambda$} \\
\hline\hline
 $0.01831..$& $0.04978..$ & $0.1353..$ & $0.3678..$  & $1$   & $2.718..$ & $7.389..$  & $20.08..$ & $54.59..$ & $148.4..$\\ 
 \hline
\end{tabular}
}
\caption{Values of $\lambda : e^{-4},e^{-3},...,e^{4},e^{5}$. }
\label{table:2}
\end{center}
\end{table}

For each value of $\lambda$ in the massless limit, we found that the optimization algorithm exhibited remarkable stable convergence to the system's minimum. 
When physical properties are plotted as functions of $\lambda$, they show remarkable agreement with their predicted scaling dependence. We first present these results for large $N$ planar quantities, and then for the spectrum of the theory.

\subsection{Planar limit}

Table \ref{table:3} displays a subset of the results obtained for the planar limit of the quantum mechanical system: the large $N$ ground state energy and the non-zero expectation values of all loops with $4$ matrices or less. We list the ground state energies with $5$ decimal places, and loop data with $4$ decimal places, as to this accuracy loops odd under $1 \to -1$ , $2 \to -2$ and $3 \to -3$ vanish and are not displayed. Symmetry under the $O(3)$ permutation subgroup is seen to be realised to at least three significant digits. 

\begin{table}[h!]
\begin{center}
\resizebox{1.1\textwidth}{!}{%
\begin{tabular}{||c|c|c|c|c|c|c|c|c|c|c||} 
\hline 
\hline
$\lambda$ & $e^{-4}$ & $ e^{-3}$ & $e^{-2}$& $e^{-1}$ & $1$ & $e^{1}$ & $e^2$& $e^3$ & $e^4$ & $e^5$ \\
\hline 
\hline
$e_{0}/N^{2}$ & 0.36048	&0.50309	&0.70211&	0.97988&	1.36753	&1.90854	&2.66359	&3.71733 &	5.18795&	7.24037\\
\hline 
$\tr 1\, /N^{2}$ &  1.0000 & 1.0000 & 1.0000 & 1.0000 & 1.0000  & 1.0000 & 1.0000 & 1.0000 & 1.0000 & 1.0000\tabularnewline
\hline 
$[11]\, /N^{2}$ &  1.6629	&1.1915	&0.8538	&0.6117	&0.4383	&0.3141	&0.2250	&0.1613	&0.1155	&0.0828\tabularnewline
\hline 
$[22]\, /N^{2}$ & 1.6628	&1.1914	&0.8537	&0.6117	&0.4383	&0.3141	&0.2250	&0.1612	&0.1155	&0.0828\tabularnewline
\hline 
$[33]\, /N^{2}$ & 1.6630	&1.1916	&0.8538	&0.6118	&0.4384	&0.3141	&0.2251	&0.1613	&0.1156	&0.0828\tabularnewline
\hline 
$[1111]\, /N^{3}$ &  5.5732	&2.8614	&1.4691&	0.7542	&0.3872	&0.1988	&0.1021	&0.0524	&0.0269	&0.0138\tabularnewline
\hline 
$[1122]\, /N^{3}$ &  2.5866&	1.3280&	0.6818&	0.3501&	0.1797&	0.0923&	0.0474&	0.0243&	0.0125	&0.0064\tabularnewline
\hline 
$[1133]\, /N^{3}$ &2.5860	&1.3277	&0.6817	&0.3500	&0.1797	&0.0923	&0.0474	&0.0243	&0.0125	&0.0064\tabularnewline
\hline 
$[1212]\, /N^{3}$ & 0.3997	&0.2052	&0.1054	&0.0541	&0.0278	&0.0143	&0.0073	&0.0038	&0.0019	&0.0010\tabularnewline
\hline 
$[1313]\, /N^{3}$ &  0.3995	&0.2051	&0.1053	&0.0541	&0.0278	&0.0143	&0.0073	&0.0038	&0.0019	&0.0010\tabularnewline
\hline 
$[2222]\, /N^{3}$ & 5.5730	&2.8613	&1.4690	&0.7542	&0.3872	&0.1988	&0.1021	&0.0524	&0.0269	&0.0138\tabularnewline
\hline 
$[2233]\, /N^{3}$ & 2.5866	&1.3280	&0.6818	&0.3501	&0.1797	&0.0923	&0.0474	&0.0243	&0.0125	&0.0064\tabularnewline
\hline 
$[2323]\, /N^{3}$ &  0.3996	&0.2052	&0.1053	&0.0541	&0.0278	&0.0143	&0.0073	&0.0038	&0.0019	&0.0010\tabularnewline
\hline 
$[3333]\, /N^{3}$ &  5.5740	&2.8618	&1.4693	&0.7544	&0.3873	&0.1988	&0.1021	&0.0524	&0.0269	&0.0138\tabularnewline
\hline 
\hline 
\end{tabular}}
\caption{Planar energies and non-zero correlator expectation values for loops with up to four matrices.}
\label{table:3}
\end{center}
\end{table}

If one fits the logarithmic plot of the large $N$ ground state energies versus $\lambda$, one finds remarkable agreement with the scaling behaviour\footnote{The parameters and their uncertainties are obtained with the Mathematica functions LinearModelFit and NonlinearModelFit.}:
\begin{equation*}
\ln e_0 / N^2 = 0.313006 + 0.33333332(1)  \ln \lambda \, .
\end{equation*}

\begin{figure}[h!]
    \centering
    \subfloat[Linear fit of $\ln e_0 /N^2$ versus $\ln \lambda$]{{\includegraphics[width=0.45\textwidth]{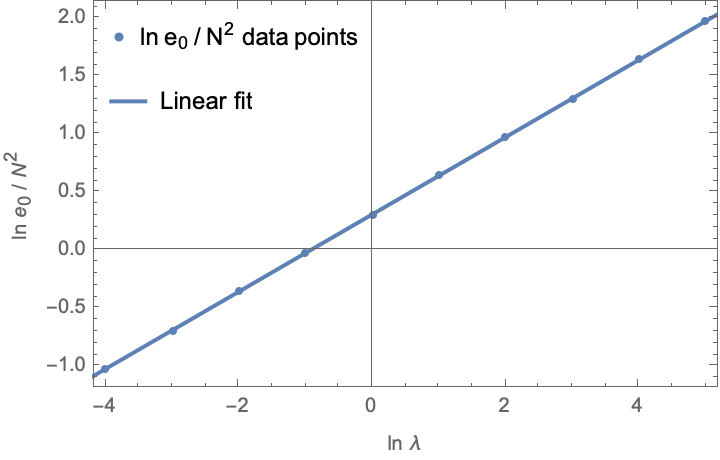} }}
    \qquad
    \subfloat[Fit of $e_0 /N^2$ to the scaling function $ 1.36753 \, \, {\lambda}^{1/3} $ ]{{\includegraphics[width=0.45\textwidth]{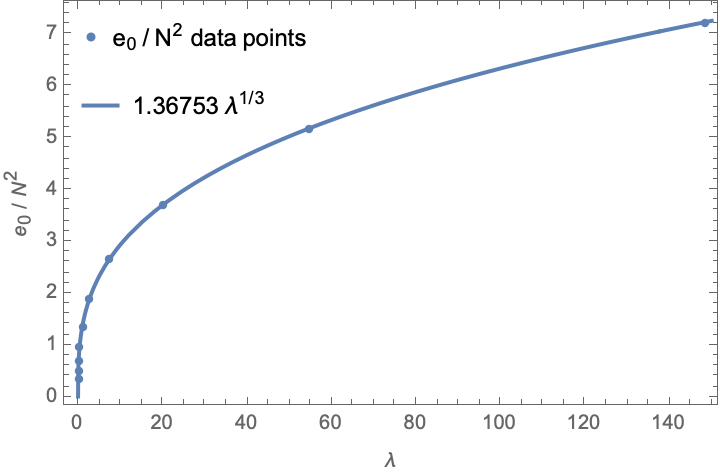} }}
    \caption{Logarithmic linear fit and fit to predicted scaling dependence for the planar ground state energy $e_0 /N^2 $. }
    \label{fig:1-E_0-vs-lambda}
\end{figure}

The accuracy with which the interpolation matches the exact scaling $p=1/3$ over of a range of couplings close to four orders of magnitude, at this level of truncation, is indeed remarkable . We are then justified in setting $p=1/3$ and fit the data to the scaling function $e_0 / N^2 = \Lambda_0 \,  \lambda^{1/3}$ with result 
\begin{equation}\label{E0ScPar}
\Lambda_0=1.36752999(4).
\end{equation}
The logarithm linear fit and the final data fit to the predicted scaling dependence are displayed in figure (\ref{fig:1-E_0-vs-lambda})

While these results are impressive, and demonstrate that the truncation scheme preserves the scaling properties of the system, errors are mainly associated with the size of the truncation (\cite{Mathaba:2023non}). In our case, this is difficult to obtain directly: the next sized truncation is not computationally feasible, and the smaller size is too small. For planar properties, we base our estimate on the accuracy with which planar quantities are displayed in Table \ref{table:3}. We then list the final scaling dependence on 't Hooft's coupling for the planar ground state energy of the massless system as:
\begin{equation*}
\boxed{
e_0 / N^2=1.3675(1) \,  \lambda^{1/3}  
}
\end{equation*}

\begin{figure}[h!]
    \centering
    \subfloat[Linear fit of $\ln \tr\,AA$ versus $\ln {\lambda}$]{{\includegraphics[width=0.45\textwidth]{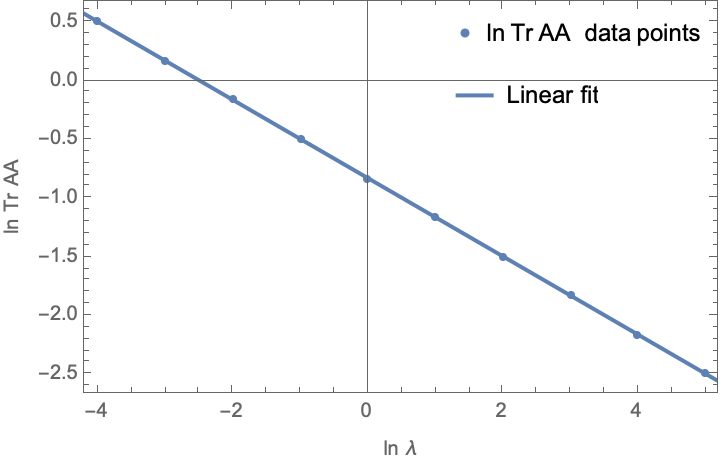} }}
    \qquad
    \subfloat[Fit of $\tr\,AA$ to scaling function $ 0.4383 \, \, {\lambda}^{-1/3} $ ]{{\includegraphics[width=0.45\textwidth]{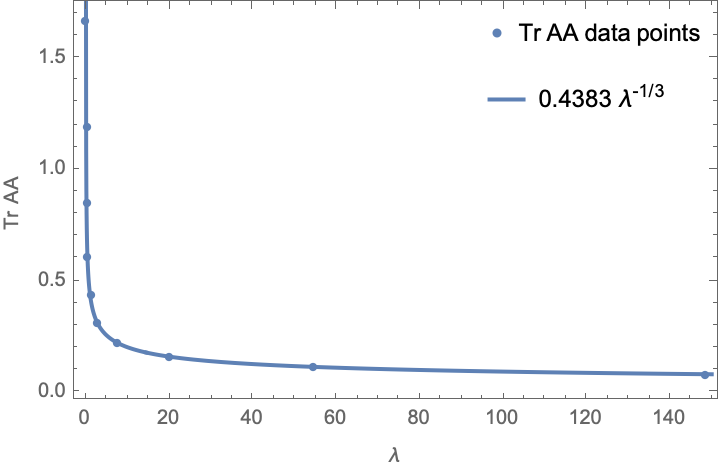} }}
    \caption{Numerical results for the planar limit of $\tr\,AA \equiv (\tr X_A X_A)/3N^2$, logarithmic linear fit and fit to predicted scaling dependence. }
    \label{fig:4-AA-vs-lambda}
\end{figure}

 We follow the same analysis for loops containing two matrices and consider the correlator $\tr\,AA \equiv (\tr X_1^2 + \tr X_2^2 + \tr X_3^2)/3N^2$. As was the case for the ground state energy, the logarithmic $\lambda$ dependence is first approximated by a linear fit $\tr\,AA = {\rm{C}}_{AA} \,  {\lambda}^p \,$ and then fitted to the scaling dimensions of the loop correlator $\tr\,AA = \Lambda_{AA} \,  {\lambda}^{-1/3}$
The results are presented in Table \ref{table:4} and displayed in Figure \ref{fig:4-AA-vs-lambda}.
\begin{table}[h!]
\begin{center}
\begin{tabular}{|| c | c || c ||c||} 
\hline
\multicolumn{2}{||c||}{Parameters of (ln) linear fit} &{ $p = -1/3$} fixed & {Final scaling function} \\\hline\hline
$\ln {\rm{C}}_{AA}$& $p$ & $\Lambda_{AA}$ & $(\tr X_1^2 + \tr X_2^2 + \tr X_3^2)/3N^2$  \\ 
\hline
-0.8247697& -0.3333335(1)&  0.4383359(1) &     { $ 0.4383(2) \, \lambda^{-1/3}$}\\ 
 \hline
\end{tabular}
\caption{$\tr\,AA$  log linear fit parameters, scaling parameter $\Lambda_{AA}$ at this level of truncation and final scaling function }
\label{table:4}
\end{center}
\end{table}

For the truncation size considered in this communication, the scaling power for the large $N$ planar correlator is again predicted with a high level of accuracy, matching with a high level of precision the expected scaling behaviour. As was the case for the planar ground state energy, the final error estimate is based on the accuracy with which planar correlators are displayed in Table \ref{table:3} (reflecting, e.g., the accuracy with which odd planar correlators vanish and $SO(3)$ symmetry). This is consistent with the analysis of (\cite{Mathaba:2023non}), where the dependence on the truncation size can be ascertained directly.

\begin{figure}[h!]
\centering
\includegraphics[width=8cm]{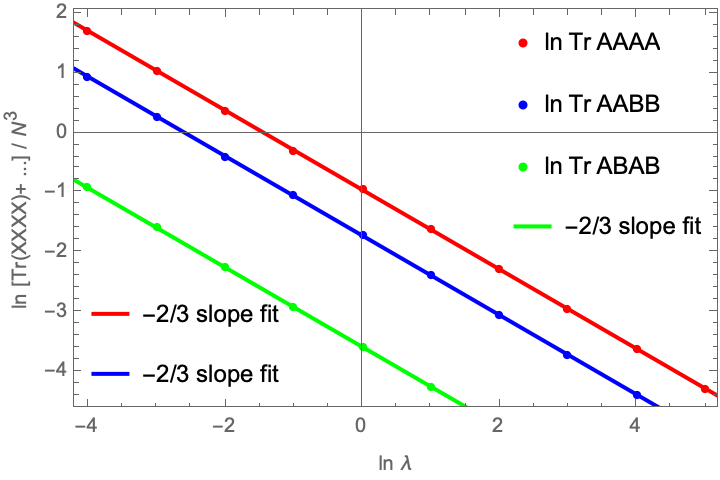}
\caption{Numerical results for the planar limits of $\tr AAAA$, $\tr AABB$ and $\tr AABB$ with logarithmic linear fits to predicted scaling dependence. }
\label{fig:5-XXXX-vs-lambda}
\end{figure}

For invariant loops with $4$ matrices, we considered the three loops 
\begin{eqnarray*}
\tr AAAA &\equiv& ([1111]+[2222]+[3333])/3 N^3 \\
\tr AABB &\equiv&([1122]+[1133]+[2233])/3 N^3 \\
\tr ABAB &\equiv&([1212]+[1313]+[2323])/3 N^3
\end{eqnarray*}
and carried out the same analysis, which is summarized in table \ref{table:5} and figure \ref{fig:5-XXXX-vs-lambda}, 
where the final logarithmic linear fits with scaling $p=-2/3$ are shown. 
\begin{table}[h!]
\begin{center}
\begin{tabular}{||c||c|c||c||c||} 
\hline
& \multicolumn{2}{|c||}{Log linear fit} & {$p=-2/3$} & Final\\
\hline\hline
 &$\ln {C}$& $p$ & $\Lambda$ & Scaling function \\ 
\hline
$\tr AAAA$  &-0.9486660 & -0.6666669(1) &   0.3872572(1)  & $0.3873(5) \, \lambda^{-2/3}$ \\ 
 \hline
 $\tr AABB$ &-1.716396& -0.6666673(1) &    0.1797126(1) & $0.1797(1) \, \lambda^{-2/3}$ \\ 
 \hline
 $\tr ABAB$ &-3.58398  & -0.666671(1)  &    0.0277649(1) & $0.0278(1)\, \lambda^{-2/3}$ \\ 
\hline
\end{tabular}
\caption{Logarithmic linear fit parameters and scaling parameter for $\tr AAAA$, $\tr AABB$ and $\tr ABAB$. }
\label{table:5}
\end{center}
\end{table}


Remarks similar to those given for the previously discussed large $N$ planar quantities, which concern the high level of accuracy of the numerical results, clearly also apply to these invariant loops with $4$ matrices.     

Finally, we consider "angles" defined as  
\begin{equation*}
\mathcal{A}_{AB} \equiv  N \frac{\tr X_A^2 X_B^2 - \tr X_AX_BX_AX_B}{\tr {X_A^2} \tr X_B^2} = -\frac{N}{2} \frac{ \tr [X_A,X_B]^2 }{\tr {X_A^2} \tr X_B^2} \, , \hspace{7pt} A \ne B \, .
\end{equation*}
Fitting constants to the loop data, one obtains: 
\begin{equation*}
\mathcal{A}_{12} = 0.790939(1) \, , \hspace{8pt} \mathcal{A}_{13} = 0.790660(1) \, , \hspace{8pt} \mathcal{A}_{23} = 0.790875(1)
\end{equation*}
Following the discussion hitherto, we assign  
\begin{equation*}
\mathcal{A} = 0.791(2) \, .
\end{equation*}
As is the case for two matrices, \cite{Mathaba:2023non}, this ratio remains constant in the massless limit for all values of the coupling constant. 
\footnote{In \cite{Han:2020bkb} the quantum mechanics of two matrices with mass was considered, and it was observed that this ratio seemed to converge to a constant value at large coupling. Its value was established directly in the massless limit in \cite{Mathaba:2023non}.} 
\footnote{For the two matrix \textit{integral}, with masses but at large coupling, it has been shown that the two matrices commute \cite{Berenstein:2008eg}, corresponding to the flat directions of the potential.}.

\subsection{$1/N$ spectrum }
We consider in this subsection the numerical results obtained for the spectrum of the theory. These are independent of $N$ and are determined from a quadratic hamiltonian $H^{(2)}_{\rm trunc}$ as $1/N$ fluctuations about the large $N$ planar background, as described in Section 2.3.

\subsubsection{Masses and scaling behaviour}
We observe that the mass of the fourth excited state and of all other higher excited states show the expected increase with coupling. The same is not the case for the three lowest lying states, which remain numerically equal to zero. These are nothing but the three commuting $U(1)$ modes of the Yang-Mills coupling, not present in a $SU(N)$ theory, as opposed to the $U(N)$ theory discussed in this communication, and as appropriate for the large $N$ limit. The fact that the numerical algorithm displays clearly identifiable physical zero modes, provides another test of the approach which we follow, as one may be concerned that the zero non-physical modes that it generates could prevent the identification of physical zero modes that a theory may possess.

Table \ref{table:6} displays the numerical results obtained for the square of the masses of the $4$th to the $37$th excited state, as a function of the coupling constant $\lambda$. Colours highlight the grouping of the states into different multiplets. Where the grouping is not entirely visible, we use information obtained from the eigenstates. This is discussed later.  

We then carry out a similar analysis to that of previous subsection, by first performing a linear fit to the log dependence of the average mass of the multiplets on the logarithm of $\lambda$, comparing it with the scaling power prediction, and then optimize the match to the scaling dependence of the energies:
\begin{equation*}
e= {A} \,  {\lambda}^p \, , \hspace{8pt} \text{and then} \hspace{5pt} e= \Lambda\, \, \lambda^{1/3} \, .\end{equation*}
This is displayed in Table \ref{table:7}, together with the quantum numbers ($J^{PC}$) of the multiplets and a final scaling function with estimated errors.  
\begin{table}
\begin{center}
\resizebox{1.1\textwidth}{!}{%
\begin{tabular}{||c|c|c|c|c|c|c|c|c|c|c||} 
\hline 
$\lambda$ & $e^{-4}$&  $e^{-3}$& $e^{-2}$& $e^{-1}$& 1 & $e^{1}$& $e^{2}$ & $e^{3}$ & $e^{4}$ & $e^{5}$\\
\hline 
\hline
\rowcolor{pink!20}
${e_4}^2$& 0.0565	&0.1101	&0.2145	&0.4180	&0.8145	&1.588	&3.091	&6.024	&11.74	&22.86\tabularnewline
\hline
\rowcolor{green!20}
${e_5}^2$& 0.2424	&0.4720	&0.9194	&1.791	&3.488	&6.794	&13.23	&25.77	&50.20	&97.77\tabularnewline
\hline
\rowcolor{green!20}
${e_6}^2$&0.2425	&0.4724	&0.9200	&1.792	&3.490	&6.798	&13.24	&25.79	&50.23	&97.84\tabularnewline
\hline
\rowcolor{green!20}
${e_7}^2$&0.2426	&0.4725	&0.9203	&1.792	&3.491	&6.800	&13.24	&25.80	&50.24	&97.86\tabularnewline
\hline
\rowcolor{green!20}
${e_8}^2$&0.2429	&0.4731	&0.9216	&1.795	&3.496	&6.810	&13.26	&25.83	&50.32	&98.00\tabularnewline
\hline
\rowcolor{green!20}
${e_9}^2$& 0.2430	&0.4733	&0.9218	&1.795	&3.497	&6.811	&13.27	&25.84	&50.33	&98.03\tabularnewline
\hline
\rowcolor{blue!10}
${e_{10}}^2$&0.5681	&1.106	&2.155	&4.196	&8.173	&15.92	&31.00	&60.38	&117.6	&229.1 \tabularnewline
\hline
\rowcolor{blue!10}
${e_{11}}^2$&0.5683	&1.107	&2.156	&4.200	&8.180	&15.93	&31.03	&60.44	&117.7	&229.3 \tabularnewline
\hline
\rowcolor{blue!10}
${e_{12}}^2$&0.5703	&1.111	&2.164	&4.214	&8.208	&15.99	&31.14	&60.65	&118.1	&230.1\tabularnewline
\hline
\rowcolor{blue!10}
${e_{13}}^2$&0.5710	&1.112	&2.166	&4.219	&8.217	&16.01	&31.17	&60.72	&118.3	&230.3\tabularnewline
\hline
\rowcolor{blue!10}
${e_{14}}^2$&0.5726	&1.115	&2.172	&4.231	&8.241	&16.05	&31.27	&60.90	&118.6	&231.0\tabularnewline
\hline
\rowcolor{blue!10}
${e_{15}}^2$&0.5740	&1.118	&2.177	&4.241	&8.260	&16.09	&31.34	&61.03	&118.9	&231.5\tabularnewline
\hline
\rowcolor{blue!10}
${e_{16}}^2$&0.5748	&1.120	&2.181	&4.247	&8.273	&16.11	&31.38	&61.13	&119.1	&231.9 \tabularnewline
\hline
\rowcolor{pink!20}
${e_{17}}^2$& 0.5947	&1.158	&2.256	&4.395	&8.560	&16.67	&32.47	&63.25	&123.2	&240.0\tabularnewline
\hline
\rowcolor{cyan!20}
${e_{18}}^2$&0.7621	&1.484	&2.891	&5.631	&10.97	&21.36	&41.60	&81.03	&157.8	&307.4 \tabularnewline
\hline
\rowcolor{cyan!20}
${e_{19}}^2$&0.7625	&1.485	&2.892	&5.634	&10.97	&21.37	&41.62	&81.07	&157.9	&307.6\tabularnewline
\hline
\rowcolor{cyan!20}
${e_{20}}^2$&0.7667	&1.493	&2.908	&5.664	&11.03	&21.49	&41.86	&81.52	&158.8	&309.3\tabularnewline
\hline
\rowcolor{yellow!20}
${e_{21}}^2$&0.9308	&1.813	&3.532	&6.879	&13.40	&26.10	&50.83	&99.01	&192.8	&375.6\tabularnewline
\hline
\rowcolor{yellow!20}
${e_{22}}^2$&0.9391	&1.829	&3.563	&6.940	&13.52	&26.33	&51.28	&99.87	&194.5	&378.9\tabularnewline
\hline
\rowcolor{yellow!20}
${e_{23}}^2$&0.9405	&1.832	&3.568	&6.950	&13.54	&26.37	&51.36	&100.0	&194.8	&379.5\tabularnewline
\hline
\rowcolor{yellow!20}
${e_{24}}^2$&0.9433	&1.837	&3.579	&6.971	&13.58	&26.45	&51.51	&100.3	&195.4	&380.6\tabularnewline
\hline
\rowcolor{yellow!20}
${e_{25}}^2$&0.9460	&1.843	&3.589	&6.990	&13.62	&26.52	&51.65	&100.6	&196.0	&381.7\tabularnewline
\hline
\rowcolor{yellow!20}
${e_{26}}^2$&0.9525	&1.855	&3.613	&7.038	&13.71	&26.70	&52.00	&101.3	&197.3	&384.2\tabularnewline
\hline
\rowcolor{yellow!20}
${e_{27}}^2$&0.9527	&1.856	&3.614	&7.040	&13.71	&26.71	&52.02	&101.3	&197.3	&384.3\tabularnewline
\hline
\rowcolor{yellow!20}
${e_{28}}^2$&0.9551	&1.860	&3.623	&7.057	&13.75	&26.77	&52.15	&101.6	&197.8	&385.3\tabularnewline
\hline
\rowcolor{yellow!20}
${e_{29}}^2$&0.9602	&1.870	&3.643	&7.096	&13.82	&26.92	&52.43	&102.1	&198.9	&387.5\tabularnewline
\hline
\rowcolor{green!20}
${e_{30}}^2$&1.229	&2.394	&4.662	&9.081	&17.69	&34.45	&67.11	&130.7	&254.6	&495.9\tabularnewline
\hline
\rowcolor{green!20}
${e_{31}}^2$&1.245	&2.425	&4.723	&9.198	&17.91	&34.89	&67.96	&132.4	&257.8	&502.2\tabularnewline
\hline
\rowcolor{green!20}
${e_{32}}^2$&1.253	&2.441	&4.755	&9.262	&18.04	&35.14	&68.44	&133.3	&259.6	&505.7\tabularnewline
\hline
\rowcolor{green!20}
${e_{33}}^2$&1.255	&2.445	&4.762	&9.274	&18.06	&35.19	&68.53	&133.5	&260.0	&506.4\tabularnewline
\hline
\rowcolor{green!20}
${e_{34}}^2$&1.263	&2.460	&4.792	&9.333	&18.18	&35.41	&68.96	&134.3	&261.6	&509.6\tabularnewline
\hline
\rowcolor{pink!20}
${e_{35}}^2$& 1.331	&2.592	&5.048	&9.832	&19.15	&37.30	&72.65	&141.5	&275.6	&536.8\tabularnewline
\hline
\rowcolor{red!10}
${e_{36}}^2$&1.485	&2.892	&5.634	&10.97	&21.37	&41.63	&81.09	&157.9	&307.6	&599.1\tabularnewline
\hline
\rowcolor{red!15}
${e_{37}}^2$&1.517	&2.954	&5.753	&11.20	&21.82	&42.51	&82.79	&161.3	&314.1	&611.8\\
\hline 
\end{tabular}}
\caption{Eigenvalues of the mass squared spectrum matrix with a truncation to $9503$ loops ($l_{\rm max} =10$) with $\Omega$ a $225 \times 225$ matrix. Only the states $n=4,...,37$ are listed.}
\label{table:6}
\end{center}
\end{table}

\begin{table}[h!]
\begin{center}
\begin{tabular}{||c|c||c|c||c||c||} 
\hline
\multicolumn{2}{||c||}{}&\multicolumn{2}{c||}{Log linear fit to mid point} & $p=1/3$ fixed& Final\\
\hline\hline
n & $J^{PC}$&$\ln {C} $& $p$ & $\Lambda$ & Scaling function  \\ 
\hline
$e_{4}$& $0^{++}$  & -0.10264 & 0.33350(2)  
&  0.9025(2)    & $0.90(2) \, \, \lambda^{1/3}$ \\ 
\hline
$e_{5...9}$ &  $2^{++}$ &0.6253110 &0.3333344(2)   &  1.868828(3) & $1.869(3)\,\, \lambda^{1/3}$ \\ 
\hline
$e_{10 ... 16}$  &$3^{-+}$&1.053395 & 0.3333291(4)& 2.86714(2) &$2.87(2)\,\, \lambda^{1/3}$  \\
\hline
$e_{17}$ &$0^{++}$ & 1.073540& 0.333340(2)&   2.92573(2)  & $2.93(3)\,\, \lambda^{1/3}$  \\
\hline
 $e_{18 ... 20}$ &$1^{-+}$& 1.198529 &0.333325(1) & 3.31522(3)   & $3.32(1)\,\, \lambda^{1/3}$  \\
\hline
 $e_{21 ... 29}$  &$4^{++}$& 1.3059729 & 0.3333382(5)  &  3.69129(2)    & $3.69(6)\,\, \lambda^{1/3}$  \\
\hline
 $e_{30 ... 34}$ &$2^{++}$&1.4445520 & 0.3333357(6) &   4.23996(2)  & $4.24(6)\, \lambda^{1/3}$ \\
\hline
 $e_{35}$ &$0^{++}$&1.476187 & 0.333325(2)&   4.37621(4)  & $4.38(3)\,\, \lambda^{1/3}$  \\
\hline
 $e_{36}$&$0^{+-}$&1.531057&  0.333340(3)& 4.62303(5) & $4.6(1)\, \lambda^{1/3}$ \\
 \hline
 $e_{37}$ &$0^{+-}$&1.54151 &  0.333325(5) &  4.67162(7)& $4.7(1)\,\, \lambda^{1/3}$  \\
 \hline
\end{tabular}
\caption{Log linear fit parameters and scaling parameters $\Lambda$ for the midpoint masses of multiplets of $n=4,...,37$ states at this level of truncation.}
\label{table:7}
\end{center}
\end{table}

Again we observe an excellent agreement with the expected scaling power $1/3$ of the coupling constant $\lambda$ for the masses of the excited bound states. Unlike the case for two matrices, where the lowest lying bound states formed a degenerate doublet with a singlet at a higher mass, for three matrices the lowest lying bound state is a singlet $0^{++}$, with the $2^{++}$ quintuplet states having highly accurate higher masses. For the singlet, we have identified some dependence on the convergence criteria, which was taken into consideration together with the "spread" of the multiplets, when applicable, in assigning error estimates to the final scaling functions displayed in Table \ref{table:7}. 

In assigning quantum numbers, $J$ is determined by the multiplicity of the multiplets. As expected, and checked with the corresponding eigenvectors,  the parity assignment is given by $(-1)^J$. The first states odd under charge conjugation are the singlets $e_{36}$ and $e_{37}$. The signature in terms of the eigenfunction components is remarkably "clean": for instance, $\Psi^{36,37}_{[1 2 3]} = - \Psi^{36,37}_{[1 3 2 ]} $,  $\Psi^{36,37}_{[1 1 2 3]} = - \Psi^{36,37}_{[1 1 3 2 ]} $ etc. to a very high degree of accuracy, with all real loop components zero with the same degree of accuracy. \footnote{$[1 2 3]^{\dagger} = [1 3 2]$, $[1 1 2 3]^{\dagger} = [1 1 3 2]$, etc.  }

\begin{figure}[h!]
    \centering
    \subfloat[Linear fit of the ln of $e_4$ and of ln of the midpoint of the $n=5,...,9$ quintuplet masses versus $\ln\lambda$ .]{{\includegraphics[width=0.45\textwidth]{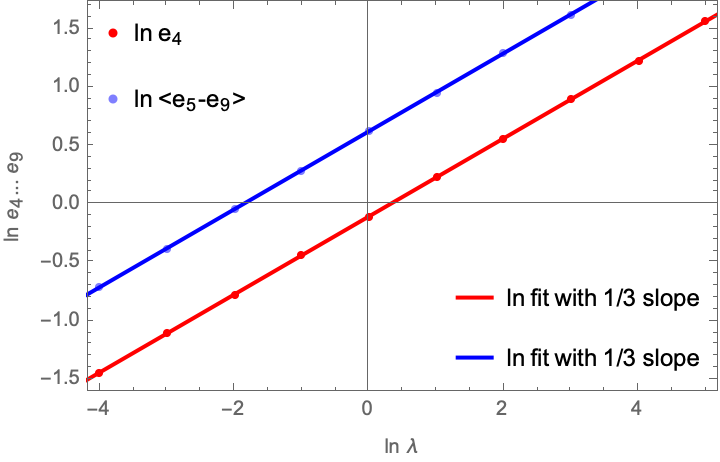} }}
    \qquad
    \subfloat[Fit of $e_4$ and of the midpoint of the $n=5,...,9$ quintuplet masses to scaling functions $ \Lambda \, \lambda^{1/3} $. ]{{\includegraphics[width=0.45\textwidth]{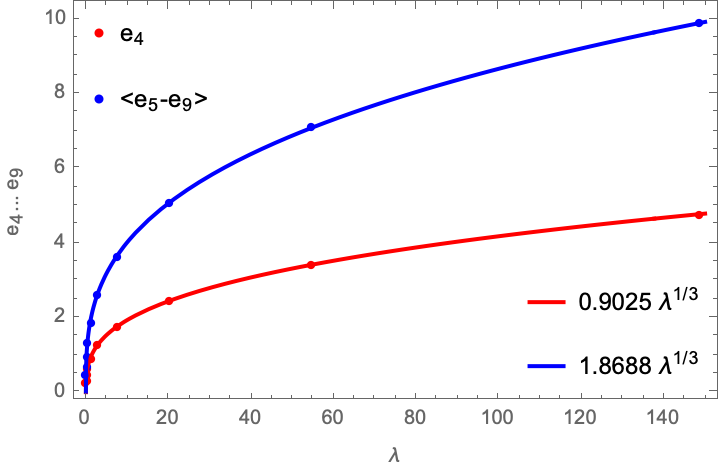} }}
    \caption{Numerical results for the mass $e_{4}$ and mid point mass of the $n=5,...,9$ quintuplet: logarithmic linear fits and fits to predicted scaling dependence.}
    \label{fig:4-2PStates-vs-g_YM}
\end{figure}

\begin{figure}[h!]
    \centering
    \subfloat[Linear fit of the ln of multiplet midpoint masses of the $n=10, ..., 20$ states versus $\ln\lambda$.]{{\includegraphics[width=0.45\textwidth]{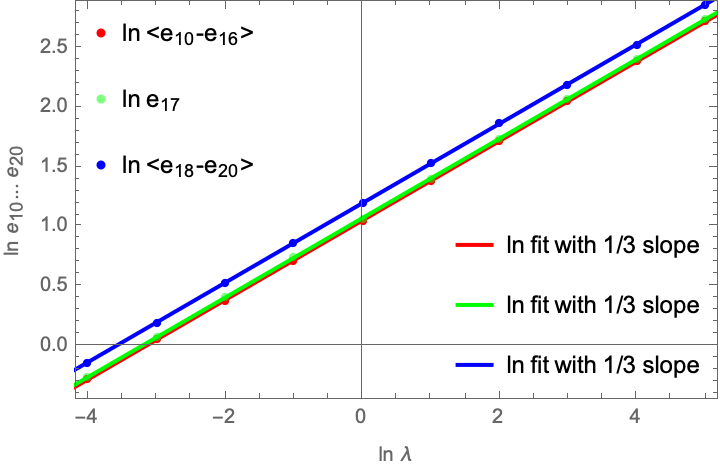} }}
    \qquad
    \subfloat[Fit of multiplet midpoint masses of the $n=10, ..., 20$ states to scaling functions $ \Lambda \, \lambda^{1/3} $ .]{{\includegraphics[width=0.45\textwidth]{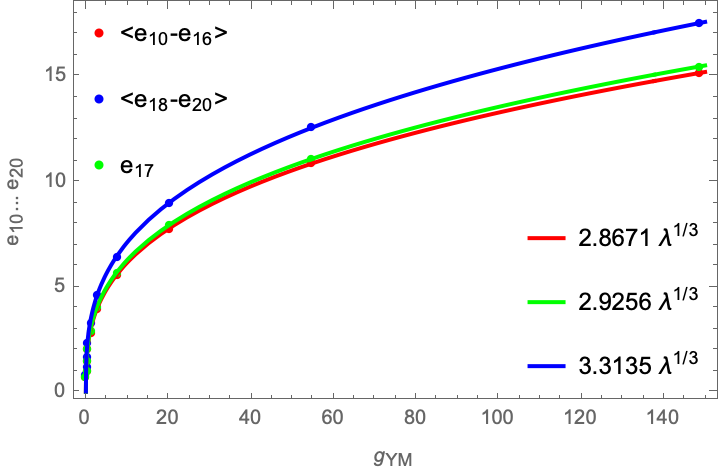} }}
    \caption{Numerical results for the multiplet midpoint masses of the $n=10, ..., 20$ states: logarithmic linear fits and fits to predicted scaling dependence.  }
    \label{fig:5-3PStates-vs-g_YM}
\end{figure}

\begin{figure}[h!]
    \centering
    \subfloat[Linear fit of the ln of the $J=4$ and $J=2$ multiplets midpoint masses of the $n=21, ..., 34$ states versus $\ln\lambda$.]{{\includegraphics[width=0.45\textwidth]{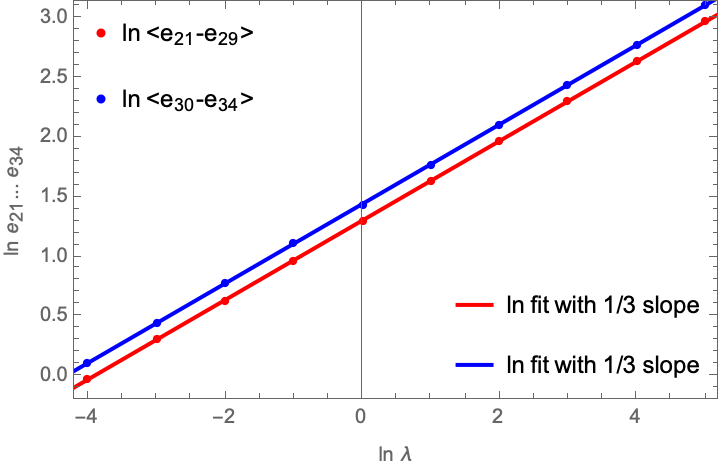} }}
    \qquad
    \subfloat[Fit of the $J=4$ and $J=2$ multiplets midpoint masses of the $n=21, ..., 34$ states to scaling functions $ \Lambda \, \lambda^{1/3} $. ]{{\includegraphics[width=0.45\textwidth]{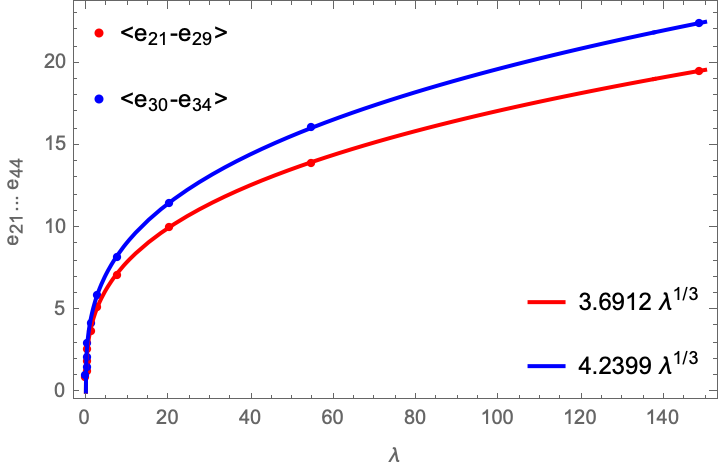} }}
    \caption{Numerical results for the $J=4$ and $J=2$ multiplets midpoint masses of the $n=21, ..., 34$ states: logarithmic linear fits and fits to predicted scaling dependence. }
    \label{fig:6-4PStates-vs-g_YM_1}
\end{figure}

\begin{figure}[h!]
    \centering
    \subfloat[Linear fit of the ln of the masses of the $n=35,36,37$ singlets versus $\ln\lambda$.]{{\includegraphics[width=0.45\textwidth]{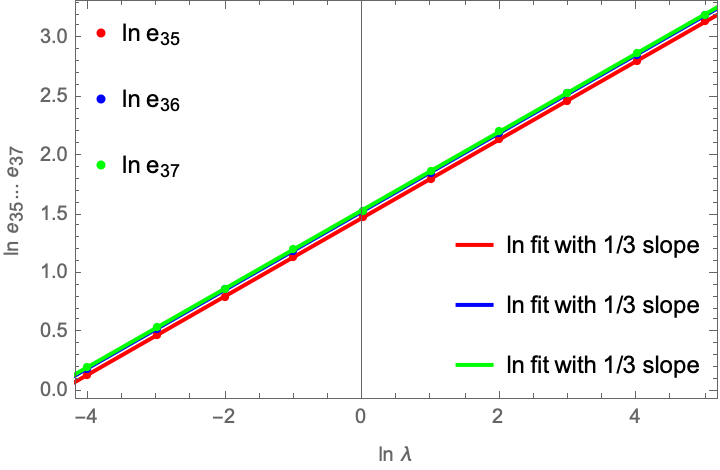} }}
    \qquad
    \subfloat[Fit of the masses of the $n=35,36,37$ singlets to scaling functions $ \Lambda \, \lambda^{1/3} $. ]{{\includegraphics[width=0.45\textwidth]{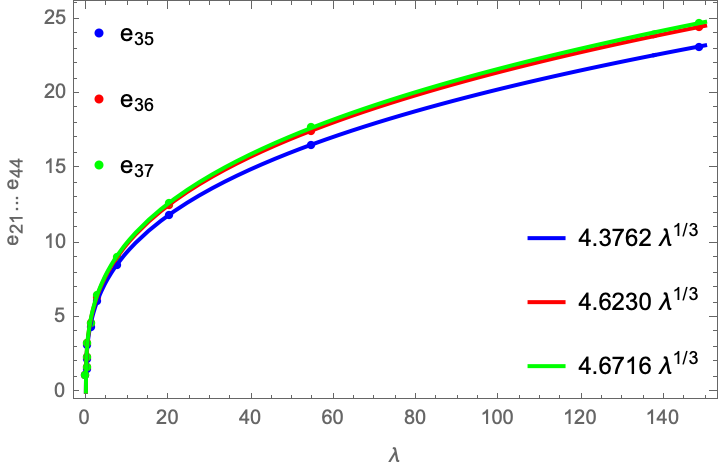} }}
    \caption{Numerical results for the $e_{35}, e_{36}, e_{37}$ singlets: logarithmic linear fits and fits to predicted scaling dependence. }
    \label{fig:7-4PStates-vs-g_YM_2}
\end{figure}

In figures \ref{fig:4-2PStates-vs-g_YM} , \ref{fig:5-3PStates-vs-g_YM},  \ref{fig:6-4PStates-vs-g_YM_1} and \ref{fig:7-4PStates-vs-g_YM_2} we display the logarithmic linear fits and the fits to the scaling power law of the numerical spectrum data for $e_n, n=4, ... , 37$. 

\section{Comparisons, discussion and outlook}
Before drawing comparisons, it is important to note that the reduced model contains states that would not be gauge invariant in the full Yang-Mills theory, which is invariant under space time gauge transformations. Examples are a mass term $\sum_{A=1}^3 \tr X_A X_A$ and the two-particle tensor. In addition, there is no particle number conservation, and as a result wave functions mix different particle number sectors. Indeed, when the eigenvectors of the mass squared matrix associated with the $0^{++}$ $e_4$ state and of the $2^{++}$ $e_ {5...9}$ states are examined, it is immediately apparent that they contain both $2$-particle and $4$-particle components, with numerically vanishing $3$-particle components\footnote{We examined the first $44$ components, and considered $\lambda=\exp{(5)}$}. 

In order to establish that spatially reduced "glueball currents" are present in the spectrum, we considered, for the $e_4$ singlet, a linear combination
\begin{equation*}
\Psi^{e_4} = \frac{a}{2} \sum_{A\ne B} [X_A,X_B][X_B,X_A] + b \big( 2 \sum_{A,B} \tr X_A X_A X_B X_B + \sum_{A,B} \tr X_A X_B X_A X_B \big).
\end{equation*}
This parametrization implies a number of consistency conditions on the loop components of the eigenvector, which are all numerically satisfied\footnote{I plan to provide more details in a forthcoming publication}. We obtain $a\sim - 0.0070$ and $b\sim -.0006$. For the $13$ component of the $2^{++}$ tensor, we took:
\begin{equation*}
\Psi^{e_6} =  a [X_1,X_2][X_2,X_3] + b \sum_{A} \big( \tr X_1 X_A X_A X_3 + \tr X_3 X_A X_A X_1 +  \tr X_1 X_A X_3 X_A \big).
\end{equation*}
Again, the loop components of the square mass matrix eigenvector satisfy numerically the consistency conditions required by this parametrization, and we obtain $a\sim-0.0115$ and $b\sim-0.010$. In both cases then, $a\ne0$, establishing that "glueball states" are present. 

We have then obtained for the ratio of the two lowest "glueball" states in the spatially reduced model:
\begin{equation*}
\frac{m_{2^{++}}}{ m_{0^{++}}} = 2.07(6) \, .
\end{equation*}
$SU(3)$ lattice gauge theory calculations, quenched (e.g., \cite{Bali:1993fb}), unquenched (e.g.\cite{Gregory:2012hu}) and as summarized in the review \cite{Vadacchino:2023vnc}\footnote{In the absence of $A_0$, there is no $O^{-+}$ in our spectrum}, give for this ratio a value of $\sim 1.5$. For a variational lattice gauge calculation extrapolated to the large $N$ limit \cite{Lucini:2010nv}, this value is increased to $\sim 1.68-1.74$, still lower than that of the reduced model result of this communication\footnote{These results include fermions in the fundamental representation. But it is known that fermion loops contributions are $1/N$ down compared to matrix valued fields loops.}. Despite this, there are expectations of volume independence in the large $N$ limit of reduced models, (e.g., \cite{Levine:1982fa, Kovtun:2007py}), albeit for unitary matrices and lattice gauge theories. Clearly, this deserves further investigation, beyond the scope of this communication.

Having in mind the importance of the "BFSS" conjecture \cite{Banks:1996vh}, the possibility of increasing the number of matrices is of great interest. Given the way in which the number of single trace invariants grow, this would seem not to be practically feasible. However, as it is apparent in this communication and in \cite{Mathaba:2023non}, we expect the large $N$ background to be invariant under the permutation subgroup of $O(d)$, with $d$ being the number of matrices. With this ansatz, the number of independent loops decreases significantly, and more matrices can be discussed. The method is easily adaptable to unitary matrices. The addition of fermionic degrees of freedom and supersymmetry are of great interest. Some of these aspects are currently under investigation.


\section{Acknowledgments }
I would like to thank Robert de Mello Koch and Mbavhalelo Mulokwe for comments on a draft of this communication.
This work is supported by the National Institute for Theoretical and Computational Sciences, NRF Grant Number 65212.

\end{document}